\documentclass[twocolumn,floats,aps]{revtex4}
\usepackage{times,amsmath,graphics,psfrag,latexsym}

\begin{document}
\title{Vortex microavalanches in superconducting Pb thin films}
\author{H.A. Radovan and R.J. Zieve}
\address{Physics Department, University of California at Davis, Davis, CA 95616}
\begin{abstract}
Local magnetization measurements on 100 nm type-II superconducting Pb thin
films show that flux penetration changes qualitatively with temperature. 
Small flux jumps at the lowest temperatures gradually increase in size, then
disappear near $T=0.7T_c$.  Comparison with other experiments suggests that
the avalanches correspond to dendritic flux protrusions.  Reproducibility of
the first flux jumps in a decreasing magnetic field indicates a role for
defect structure in determining avalanches.  We also find a
temperature-independent final magnetization after flux jumps, analogous to
the angle of repose of a sandpile. 
\end{abstract}
\maketitle

\section{Introduction}

Magnetic fields penetrate type-II superconductors in the form of vortices.
Pinning sites inside the sample maintain a spatial variation in the
vortex density, with an accompanying current density $j$. The resulting
Lorentz force drives further flux penetration.  In the Bean critical-state
model, vortex motion is triggered whenever the current density exceeds
a critical current density $j_c$, thereby maintaining a current density
$j_c$ everywhere in the material \cite{Bean}. The constant current density
corresponds to a constant vortex density gradient, much like the surface
of a sandpile with grains poised to flow. As the applied
field changes, the flux density adjusts steadily, giving rise to a smooth
hysteresis loop $B(H)$ \cite{Campbell}.

However, in many situations abrupt flux jumps appear, corresponding to
near-instantaneous motion of many vortices.  While individual moving
vortices are like single grains falling down the side of a sandpile,
flux jumps resemble an avalanche of many grains. In 1968 Heiden et
al. measured jumps 10 to 10,000 vortices in size in tubular Pb-In alloys
\cite{Heiden}.  Since then magnetic instabilities have been observed
down to 0.001 $T_c$ in YBaCuO \cite{Seidler, Zieve} and up to $0.95
T_c$ in Nb \cite{James}. Other superconductors showing flux jumps
include Nb-Ti \cite{Field},
Pb-In \cite{Stoddart}, and MgB$_2$
\cite{Zhao}.  

The most common explanation for the flux jumps is a magnetothermal
instability.  Moving flux increases the local temperature.  The
temperature change reduces $j_c$, which triggers further vortex motion.
This feedback produces the vortex avalanche.  If the sample recovers
without quenching completely, there are small steps in the magnetic
hysteresis loop rather than full jumps to zero magnetization.  Other
proposed sources for the avalanches include self-organized criticality
\cite{Richardson}, dynamical instabilities \cite{Zieve}, and stick-slip
dynamics \cite{Barford}.

Several contrasting temperature and field dependences have been
observed. Most commonly flux instabilities occur at low magnetic fields,
where the critical current density is largest, but some of the YBaCuO work
finds flux jumps only at {\em large} fields \cite{Seidler, Zieve}.  Jumps
range in size from a few vortices, requiring careful time and voltage
resolution to identify that a jump occurred, to ``complete" avalanches
that occur throughout the sample and reduce the sample magnetization
to zero.  Generally when a sample displays complete or near-complete flux
jumps the avalanche size distribution has a sharp peak, while in other
cases the distribution is broad, with either power-law or exponential
behavior.  A few recent studies have found {\em both} behaviors, fairly
small jumps with a broad size range and larger avalanches with a single
characteristic size, at different temperatures in a single sample.
In Nb samples complete flux jumps occur at the lowest investigated
temperatures and smaller avalanches at higher temperatures \cite{James,Nowak,
Behnia}, with a sharp change in jump size.  On the
other hand, in MgB$_2$ flux jumps gradually get larger as temperature
increases without ever becoming complete \cite{Zhao, Johansen01a}.

Here we report local Hall probe measurements in a new system, type-II Pb
thin films.  We find flux avalanches for temperatures below $0.7 T_c$,
with both large and small jumps at different temperatures.  The behavior
is most similar to that of MgB$_2$, with larger jumps observed only
at relatively high temperature.  We investigate sample, field history,
magnet ramp rate, and maximum cycling field dependence.  The properties
independent of external parameters are the field of occurrence and
size of the first microavalanche, and the final magnetization after
a jump.  We argue that an interplay between the vortex density and
the microstructure is at the origin of the flux instabilities. We
find that the changes in jump size with temperature arise from the
earlier triggering of avalanches at low temperature combined with the
near-constant final magnetization.

\section{Experimental}

We fabricate 100 nm Pb films on 4 mm by 4 mm Si substrates by resistive
evaporation at 2 \AA/s  at room temperature. A 40 nm Ge capping layer
prevents Pb oxidation. The thickness is calibrated with a profilometer. We
prepared the two samples discussed in this paper under nominally identical
conditions. The upper critical field was measured in a Quantum Design
MPMS XL SQUID system. For $T = 0$ K, $B_{c2}$ extrapolates to 1319 G. We
estimate the Ginzburg-Landau parameter at $\kappa = 1.23$, comfortably
within the type-II regime.

\begin{figure}
\begin{center}
\scalebox{0.9}{\includegraphics{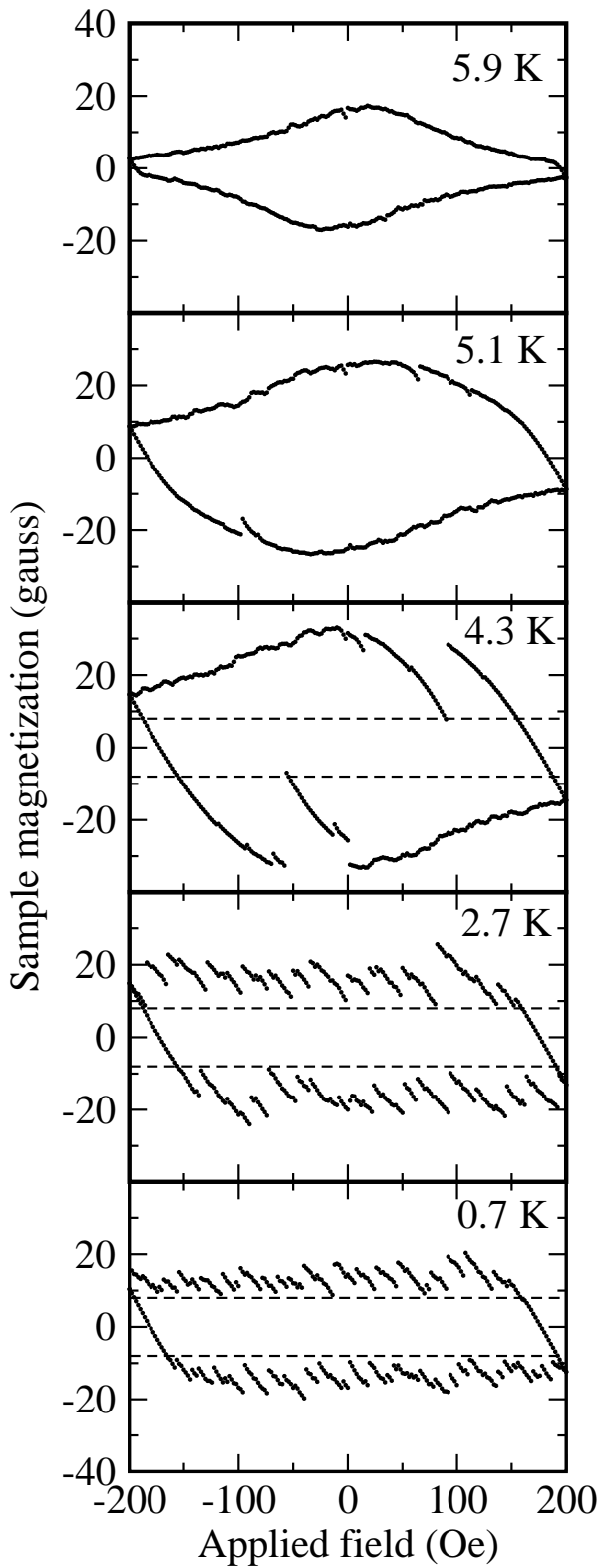}}
\caption{\small Magnetic hysteresis loops at several temperatures for Sample B. 
All graphs have the same vertical scale.  The dashed horizontal lines at $\pm 8$
gauss in each of the bottom three graphs illustrate the temperature-independence
of the magnetization just after an avalanche.}
\label{f:hystloops}
\end{center}
\end{figure}

For magnetization measurements we use a semiconductor Hall probe
\cite{AREPOC} with an active area of 400 $\mu$m$^2$ and a sensitivity
of 900 m$\Omega$/kG.  Applied magnetic fields reach 400 Oe.
Most of the measurements at $T > 4.2$ K were carried out in a simple He-dewar insert
with temperature stability better than $\pm 1$ mK.  For Sample B we
extended the data down to 0.27 K on a pumped ${}^3$He cryostat with
stability better than $\pm 10$ mK. The critical temperature for all
samples is the bulk value, $T_c = 7.2$ K. For comparison to bulk type-I
Pb we used a disk with a diameter of 4 mm and a thickness of 400 $\mu$m.

\section{Results and Discussion}

Figure \ref{f:hystloops} shows the flux jump characteristics for our
Sample B.  The maximum field of 200 Oe allows full flux penetration at
all temperatures.  The horizontal axis shows the applied, external field,
with no adjustment for demagnetization effects. The step size is 2 Oe,
with 2 seconds between steps. The width and qualitative flux jump behavior
in the hysteresis loops are fully reproducible on different cooldowns.
Just below $T_c$ the hysteresis loops are smooth on the scale of our
measurements, as represented in Figure \ref{f:hystloops} by the 5.9 K
hysteresis loop. We do not have the temporal resolution of Behnia et al.,
who show that tiny flux jumps may drive even magnetization changes that
appear smooth \cite{Behnia}. Cooling produces first a few small avalanches
(5.1 K), then larger, often nearly complete jumps which coexist with the
small ones (4.3 K). Both large and small jumps occur preferentially on
the decreasing branch of the hysteresis loop, appearing on the increasing
branch only at lower temperatures. On further cooling, the
small jumps remain, while the large ones gradually shrink until the
two types are no longer clearly distinguishable.  The hysteresis loops
also narrow and flatten as temperature decreases, with the {\em final}
magnetization just after a flux jump nearly temperature-independent.

\begin{figure}
\begin{center}
\scalebox{0.45}{\includegraphics{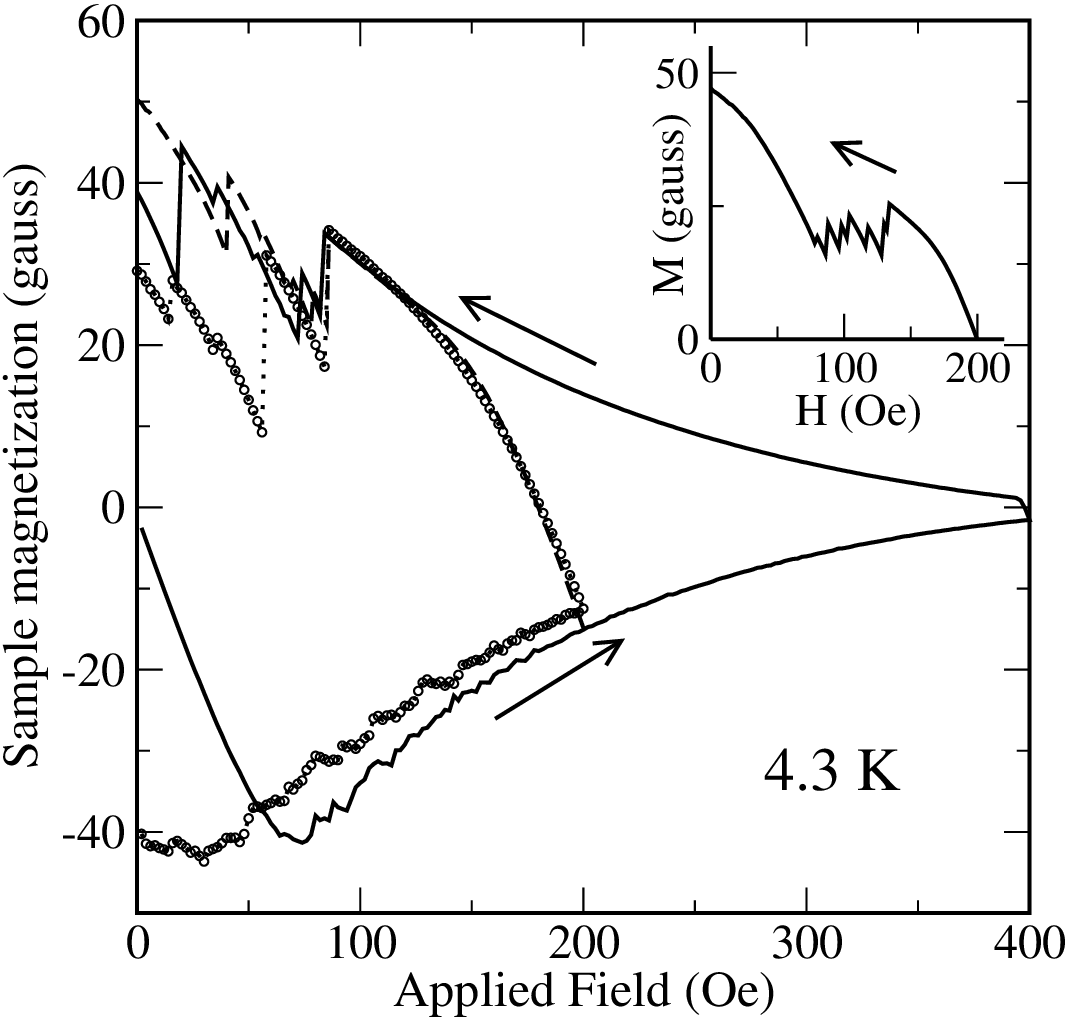}}
\caption{\small Hysteresis loops for Sample B with maximum field 200 Oe (circles
and dashed line), and 400 Oe (solid line).  Note the lack of jumps at higher
fields, and the reproducibility of the field for the first jump.  Inset: descending
branch of hysteresis loop for Sample A at 4.57 K.}
\label{f:maxH}
\end{center}
\end{figure}

The flux jumps occur only at sufficiently small fields.  The solid
hysteresis loop of Figure \ref{f:maxH} has maximum field 400 Oe, while
the dashed curve (descending branch only) begins at 200 Oe. The third
curve, represented by circles and a dotted line, was taken seven weeks
later after thermally cycling to room temperature.  It also has maximum
field 200 Oe. The hysteresis loops display some noise on the increasing
branch but no clear avalanches.  Noise on the increasing branch also
develops through the top three frames of Figure \ref{f:hystloops}.  The noisy
region, which is confined to low applied fields and increases in range
as temperature decreases, may be a precursor for the avalanches on the
increasing branch at lower temperatures.  On the decreasing branch of the
hysteresis loop, all three curves of Figure \ref{f:maxH} have their first
flux jump at the same field, regardless of the field history. This onset
field does depend on temperature, increasing as temperature decreases.
By 300 mK, the onset field for jumps exceeds 300 gauss.

Sample A, which we measured only above 4.2 K, shows generally similar
behavior. One difference is that at one temperature (4.57 K), there
is a minimum applied field for avalanches as well as a maximum field.
The inset of Figure \ref{f:maxH} illustrates this behavior after the
sample is cooled through $T_c$ in an applied field of 200 gauss; the
same effect occurs upon cooling in zero field.  Having flux jumps confined to
such a limited field range requires a particularly delicate balance
between stable and unstable regimes.  The different behavior of our two
samples also highlights the importance of the precise defect structure
in the flux jump patterns.

Among previously measured materials, MgB$_2$ behaves most like our Pb samples. 
MgB$_2$ shows flux jumps below about $t = T/T_c = 0.25$, with the jumps
steadily shrinking as $t$ decreases; and smooth changes in magnetization from
$t=0.25$ to $T_c=39 K$ \cite{Johansen01a,Zhao}.  As in our samples, the MgB$_2$
instabilities at the highest temperatures occur primarily for decreasing
magnetic field, perhaps because of an additional heat load from annihilation of
vortices with antivortices \cite{Johansen01b}.

On a microscopic level, both magneto-optical (MO) imaging and Bitter patterns
in a variety of materials show that many flux jumps come from sudden
dendritic protrusions of high-vortex-density regions into the specimen
\cite{James,Johansen01a,Johansen01b,Runge,Duran, Leiderer}. In MgB$_2$, MO
shows quasi-one-dimensional dendritic fingers at the lowest temperatures, and
branched structures at higher temperatures, with a similar behavior found in
computational work \cite{Johansen01a, Aranson}. Although MO and magnetization
measurements have rarely been carried out on the same sample, the slight
increase in avalanche size with temperature in MgB$_2$ apparently corresponds
to the increased branching of the dendrites, which allows more flux motion
\cite{Johansen01a}. 
In Nb, however, MO shows dendritic ``fingers" at both low
\cite{Heiden} and intermediate \cite{James} temperatures, while magnetization
measurements consistently find the largest avalanches at low temperatures. 
We suggest that the complete flux jumps found in Nb at the lowest
temperatures differ fundamentally from the partial jumps in MgB$_2$. Indeed,
at low temperatures the flux front in Nb moves in spurts but without
branching \cite{James}. With no optical data on Pb, we cannot firmly identify
our flux jumps as coming from fingering events, but similarities between the
jumps in our samples and in MgB$_2$ make this a good possibility.  The
biggest difference is the quantitative observation that our high-temperature
avalanches, while not complete, are substantially larger than those in
MgB$_2$.

We have eliminated various possible artifacts as the source of our
results. First, with no sample or with a bulk Pb disk, the Hall probe
yields no fluctuations in the signal down to the lowest
temperatures. Hall probe excitation currents from 5 $\mu$A to 1 mA at $T =
0.3$ K do not change the loop width or jump sizes, so heat from the Hall
current is not a factor within this range.  Increasing the excitation
current to 5 mA, however, does reduce the width notably. The avalanches
are also independent of ramp rates from 0.2 Oe/s to 3.3 Oe/s, data point
spacings from 1 Oe to 10 Oe, and history effects such as the maximum field
achieved during a hysteresis loop or cooling in zero or non-zero field.

\begin{figure}
\begin{center}
\scalebox{0.45}{\includegraphics{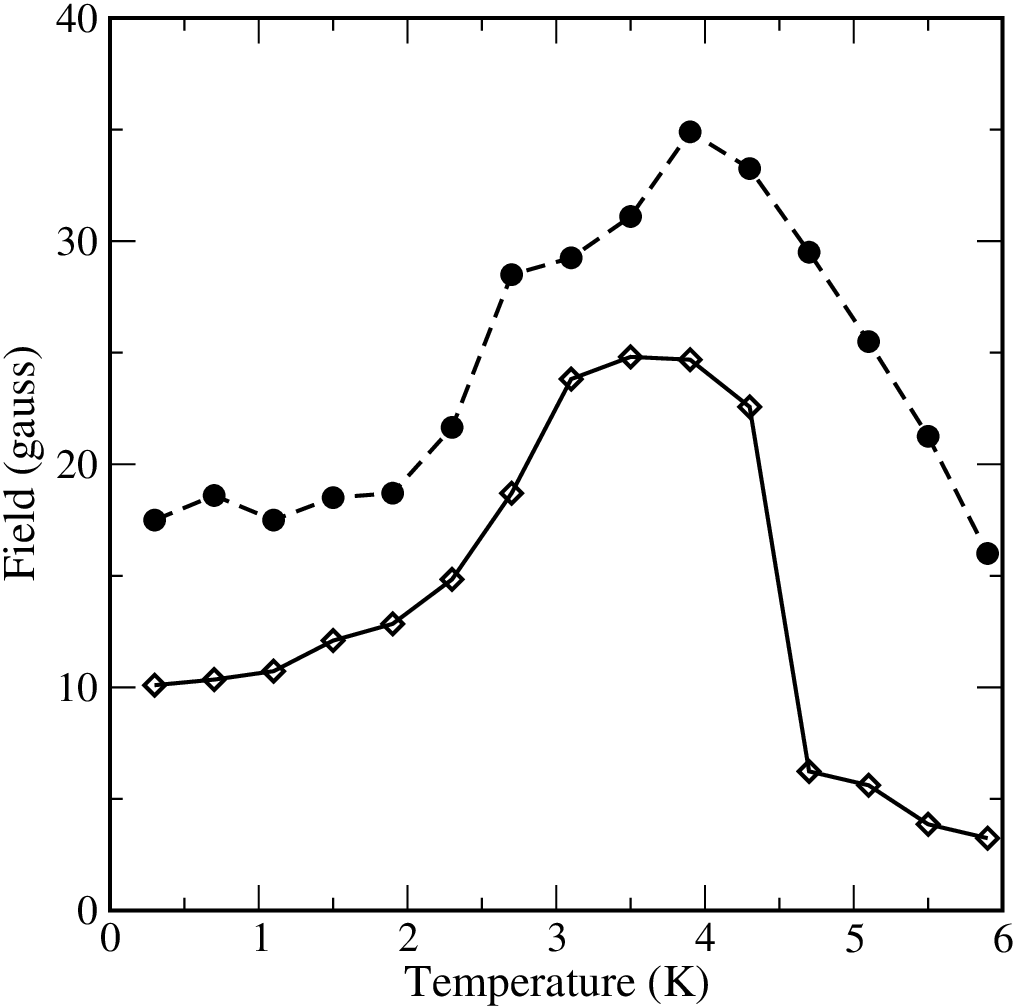}}
\caption{\small Half the maximum width of hysteresis loop (filled circles) and
size of second-largest avalanche (diamonds).  The two curves track each other
below 3 K.}
\label{f:width}
\end{center}
\end{figure}

Figure \ref{f:width} displays the maximum magnetization as a function
of temperature for Sample B.   The values used are half the maximum
difference between the ascending and descending branches of a hysteresis
loop.  In critical-state models, the hysteresis loop width is directly
proportional to the critical current density $j_c$ of the material and
increases steadily as temperature drops. We find this behavior
above 3.9 K, but on further decreasing temperature the magnetization
drops rapidly. The decrease becomes less steep but persists down to
our lowest temperature of 0.27 K, where the width is 60\% below its
maximum value.  This narrowing of the hysteresis loop is also visible
in Figure \ref{f:hystloops}.  Once the flux jumps begin, the idea of
a critical state may no longer apply.  The numerous low temperature
avalanches significantly depress the current carrying ability of
type-II Pb films. Although vortex avalanches are often attributed to
thermomagnetic instabilities, the reduction of the loop width shows that
achieving a global critical current is not a requirement for triggering a
jump. Qualitatively similar behavior was recently found in MgB$_2$ films,
where $j_c$ drops up to 40\% below $t = 0.25$ \cite{Johansen01a,Zhao}.

Along with the maximum magnetization, we plot the size of the
second-largest avalanche at each temperature.  We choose the
second-largest rather than the largest avalanche because there is less
variation in size, although using the largest avalanche gives similar results. The
jump size at low temperature decreases roughly as $T^3$, approaching
a finite value as $T\rightarrow 0$.  Significantly, the loop width
and avalanche size track each other closely, indicating that the final
magnetization is nearly independent of temperature.  Again, this point is
illustrated directly in Figure \ref{f:hystloops}, where the dashed lines
at $\pm 8$ gauss for the bottom three frames show how near to this field
the avalanches end. The final magnetization does vary between cooldowns,
and was about 12 gauss and 19 gauss for two other cooldowns on Sample B.
The variation may stem from the heat-sinking of the substrate or from changes
in the location of the Hall probe.

Our measurements show that the flux jump trigger changes with
temperature. At the higher temperatures, the data are consistent with
thermomagnetic instabilities within a Bean-type critical state model,
since jumps begin only when the magnetization lies along the ideal
hysteresis loop. At lower temperatures, the narrowing hysteresis loops
show that the avalanches begin before the sample reaches a global
critical state.  The nearly constant envelope of the magnetization with
applied field shows that at low temperatures the flux jump trigger
becomes independent of field.  The narrowing of the hysteresis loop
begins at about the same temperature where flux jumps start to occur for
increasing field.  If the jumps correspond to dendrites of flux entering
the sample, the dendrites themselves could produce local field
variations that trigger further avalanches while the sample is well away
from a global critical state.

Furthermore, the uniform final magnetization shows that the cessation
of the avalanches has a different mechanism from their onset, and
is independent of whether the initial trigger is global or local.
The temperature-independence of the final magnetization suggests that
the jumps do not halt simply from a thermal recovery.  Since the lowest
temperature flux jumps are also the smallest, the system should not reach
as high a temperature, leaving no clear origin for the constant
final magnetization.  Rather, the moving vortex system seems to recover
upon reaching a particular current density which acts much like the
angle of repose of a sandpile.  Once again our sample behaves
much like MgB$_2$, where recent measurements
find that the local field just after an avalanche has a
reproducible maximum value of about 120 gauss \cite{Barkov}.

\begin{figure}
\begin{center}
\hspace*{-.2in}
\scalebox{0.55}{\includegraphics{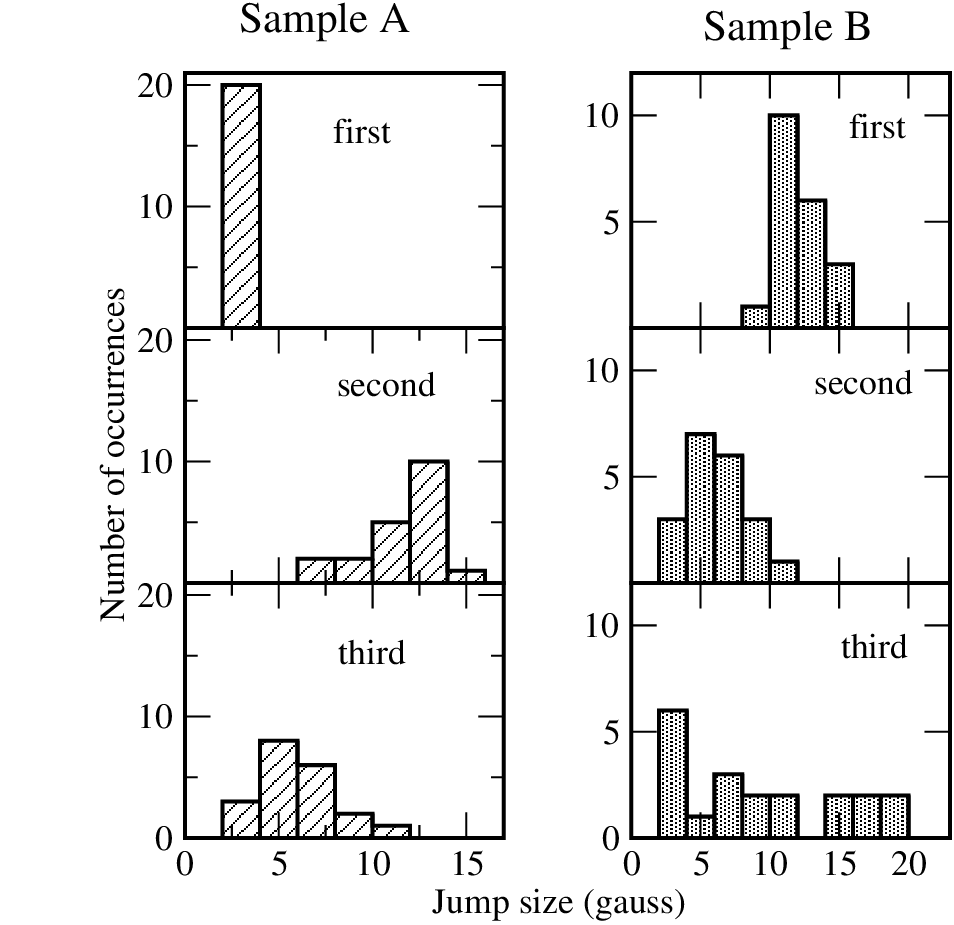}}
\caption{\small Size distributions for first three avalanches for
Sample A at 4.35 K and Sample B at 4.6 K.}
\label{f:firstav}
\end{center}
\end{figure}

As noted above, the three loops of Figure \ref{f:maxH} have their first
flux jumps at nearly identical fields.  In fact both the size and the
applied field of the first avalanche are robust against changes in ramp
rate, maximum cycling field, and field history, although they are sample
dependent.  However, the sharp peaks in size disappear quickly. Figure
\ref{f:firstav} shows histograms for the first three avalanches.
The data for each sample come from a series of 20 identical half-loops
with ramp rate 1 Oe/s on the decreasing branch.  The field and avalanche
size for the first jump is $H =$ (130 $\pm$ 1) Oe with $\Delta B =$ 2 -
3 G for Sample A and $H =$ (84 $\pm$ 3) Oe with $\Delta B =$ 10 - 15 G
for Sample B. The parameters are particularly repeatable for Sample A,
but even the distribution in Sample B is narrower than that of later
avalanches. The second avalanche is broader than the first but still
has a characteristic size.  The third and subsequent avalanches follow
a broader distribution weighted towards small sizes.  As temperature
decreases, the distinction between the early and later avalanches goes
away, disappearing entirely by 2.7 K.  The reproducibility of the field
for the first avalanche vanishes likewise.  A possible interpretation
is that the initial flux jumps from a smooth flux front are primarily
influenced by defects in the sample.  The similarities of the first few
jumps on different hysteresis loops reflect the nearly identical flux
penetration. Small irreproducibilities in the first few jumps leave
unique flux profiles. These magnetization patterns, as well as the
defect structure, influence later flux jumps, destroying further
quantitative likeness among hysteresis loops.  The characteristic
initial jump size disappears at low temperature because once
flux jumps occur for increasing field, even the initial magnetization
pattern on the decreasing branch varies among loops.

\begin{figure}
\begin{center}
\psfrag{1}{\scalebox{2}{$10^0$}}
\psfrag{10}{\scalebox{2}{$10^{1}$}}
\scalebox{0.6}{\includegraphics{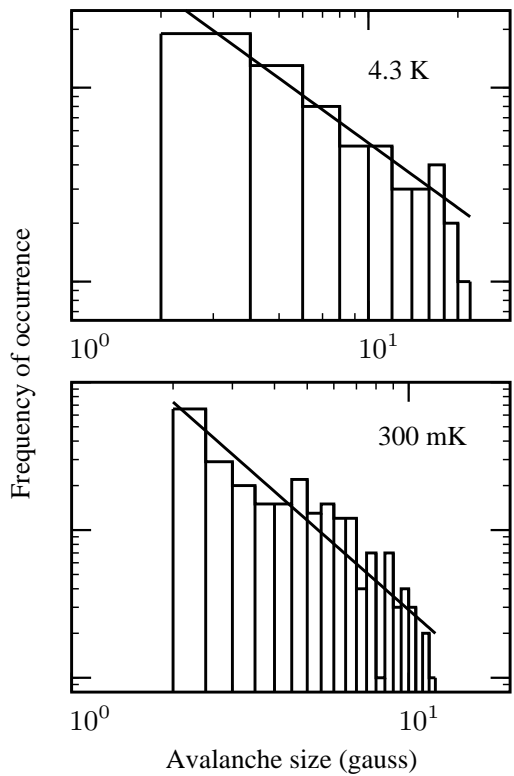}}
\caption{\small Size distributions for Sample B at 4.3 K (top) and 300 mK
(bottom), on log-log scales.  The first two avalanches per loop are omitted from
the 4.3 K data. Solid lines are power law fits, with exponents -1.09 for 4.3 K and
-2.01 for 300 mK.} 
\label{f:histogram}
\end{center}
\end{figure}

For both our samples the individual avalanches range in size from
about 1 G to 16 G, corresponding to a change of 20 to 300 vortices
under the Hall probe. The similar size range indicates that the same
general mechanism is responsible for these flux front instabilities.
Figure \ref{f:histogram} shows the size distribution of flux jumps for
Sample B during 20 cycles at $T = 4.3$ K, where flux jumps occur only
for decreasing field, and three loops at $T=300$ mK with flux jumps on
both branches.  We omit the first two flux jumps at 4.3 K because of their
atypical size distribution, as discussed above.  Interestingly, at 4.3 K
the avalanche sizes do {\em not} actually have a sharp division between
large and small; it appears so for an individual hysteresis loop only because there are
so few large jumps. Both exponential \cite{Duran,Heiden} and power-law
\cite{Nowak,James} distributions are reported in the literature. For
our data a power law form works much better than an exponential for
the 300 mK data, and somewhat better at 4.3 K. The best-fit powers for
these two temperatures are very close to -2 and -1, respectively. The
special character of the first two avalanches may account for some of
the controversy over whether the size distribution follows power law
or exponential behavior.  Including them at 4.3 K  changes the best
fit from power law to exponential, although neither function fits the
distribution especially well.

Finally, since the repeatability of our initial avalanches suggests that
defect patterns influence the flux jumps, we comment on microstructure.
Our Pb thin films experience significant extrinsic stress from their
adhesion to the Si substrate. The thermal expansion coefficients of
Si and Pb differ by one order of magnitude. Upon thermal cycling Pb
releases compressive stress by atomic diffusion, which forms hillocks
and voids. Another source of stress is the 8.9\% lattice misfit between
film and substrate, and the 12.4\% misfit between the Pb film and the
Ge capping layer. The film releases this stress through formation of
predominantly edge-type dislocations \cite{Tu}. Any two samples have
different defect structures, and the existence of a minimum field for
avalanches in only one of our samples shows that details of the defects
can strongly affect flux stability.   We have also tested a bulk Pb
disk, which is high purity and has no misfit stress.  The disk does not
support flux instabilities in any temperature or field range, during
flux entry or exit.  Although bulk lead is a type-I superconductor,
MO investigations show that field penetration begins with flux tubes
containing 500 to 1000 flux quanta \cite{Kirchner}.  The flux tubes
form a hexagonal lattice much like the Abrikosov vortex lattice, and
in principle should be able to undergo avalanching similar to that of
quantized vortices in type-II materials. The absence of microavalanches
in the bulk material is consistent with its relative scarcity of defects.

\section{Summary}

We report local magnetic measurements on 100 nm Pb type-II thin films for
temperatures down to 0.27 K. The hysteresis loops display several flux
penetration patterns as a function of temperature, starting out with many
microavalanches at the lowest temperature, then fewer and bigger ones
until the classical critical-state type flux penetration is reached for
$T/T_c > 0.7$.  We draw attention to two surprisingly robust features:
the size and location of  the first instability in a decreasing magnetic
field, and the final magnetization after an avalanche. The occurrence
of the first instability varies little with external parameters,
but is sample dependent.  The final magnetization also varies among
cooldowns, but is nearly temperature-independent for a given cooldown
from room temperature.  Finally, we note that the similarity of our
work and recent measurements on MgB$_2$ films shows that the underlying
mechanisms governing vortex motion are not specific to MgB$_2$.

\section{Acknowledgement}

We thank P. Klavins for technical assistance.
This work was supported by the NSF under DMR-9733898.

\end{document}